\documentclass[conference,11pt]{IEEEtran}
\usepackage{cite}
\usepackage{amsmath,amssymb,amsfonts}
\usepackage{algorithmic}
\usepackage{graphicx}
\usepackage{textcomp}
\usepackage{xcolor}
\usepackage{multirow}
\usepackage{csquotes}
\usepackage{subcaption} 
\usepackage{caption}
\usepackage{microtype}
\usepackage{booktabs}

\usepackage{tikz}
\usetikzlibrary{arrows.meta,positioning,fit,shapes.misc,shapes.symbols}

\def\BibTeX{{\rm B\kern-.05em{\sc i\kern-.025em b}\kern-.08em
    T\kern-.1667em\lower.7ex\hbox{E}\kern-.125emX}}
    
\begin{document}

\title{When the Base Station Flies: Rethinking \\ Security for UAV-Based 6G Networks}

\author{
    \IEEEauthorblockN{Ammar El Falou}
\IEEEauthorblockA{
Computer, Electrical and Mathematical Sciences and Engineering (CEMSE) Division \\ 
King Abdullah University of Science and Technology (KAUST) \\
Thuwal, Saudi Arabia\\}
Email: ammar.falou@kaust.edu.sa}

\maketitle


\begin{abstract}
The integration of non-terrestrial networks (NTNs) into 6G systems is crucial for achieving seamless global coverage, particularly in underserved and disaster-prone regions. Among NTN platforms, unmanned aerial vehicles (UAVs) are especially promising due to their rapid deployability. However, this shift from fixed, wired base stations (BSs) to mobile, wireless, energy-constrained UAV-BSs introduces unique security challenges. Their central role in emergency communications makes them attractive candidates for emergency alert spoofing. Their limited computing and energy resources make them more vulnerable to denial-of-service (DoS) attacks, and their dependence on wireless backhaul links and GNSS navigation exposes them to jamming, interception, and spoofing. Furthermore, UAV mobility opens new attack vectors such as malicious handover manipulation. This paper identifies several attack surfaces of UAV-BS systems and outlines principles for mitigating their threats. 

\end{abstract}

\vspace{6pt}

\section{Introduction}
The integration of non-terrestrial networks (NTNs) into 5G-Advanced and 6G systems is a key enabler for global connectivity, especially in underserved and disaster-prone regions \cite{dang_what_2020,kaltenberger_driving_2025,wang_unmanned_2024,bajracharya_6g_2022,el2025study}. While terrestrial networks (TNs) provide good connectivity in urban and suburban areas. Still, they often fail to provide coverage in rural areas, during disasters, and in mega-sized events \cite{matracia_uav-aided_2023,shehab_five_2024,ben_salem_exploiting_2023}. The 3rd Generation Partnership Project (3GPP) defines NTN as network segments that utilize an airborne or spaceborne vehicle for transmission, such as satellites, high-altitude platform systems (HAPS), and unmanned aerial vehicles (UAVs) \cite{hassan_ntn_2023,pugliese_integrating_2024}. NTNs extend the reach and usability of cellular networks far beyond the limitations of terrestrial infrastructure. Since Release 15, 3GPP has progressively incorporated NTN features, with 5G-Advanced (Release 18) enabling NTN-specific enhancements and 6G envisioned as a fully integrated space-air-ground network with seamless handovers between TN and NTN.

Within this vision, UAVs complement terrestrial networks (TNs), HAPS, and satellites to form a highly scalable communications infrastructure. Unlike satellites or HAPS, UAVs can be rapidly deployed, making them particularly valuable for disaster recovery, temporary capacity boosts, and rural coverage \cite{shehab_five_2024,tharakan_efficient_2024}. For UAVs acting as base stations (UAV-BS), the requirements are simpler than those for other types of NTN. The wired backhaul link is to be replaced with a wireless link, and the UAVs should operate on the access link according to the 3GPP standard \cite{mundlamuri_integrated_2023}.  

\textbf{Position.} When the BS flies, the security challenges change. We aim to: (i) expose critical vulnerabilities for UAV-based NTNs, and (ii) outline mitigation techniques for a secure 6G UAV-based architecture.

\begin{figure}
    \centering
    \includegraphics[width=0.95\linewidth]{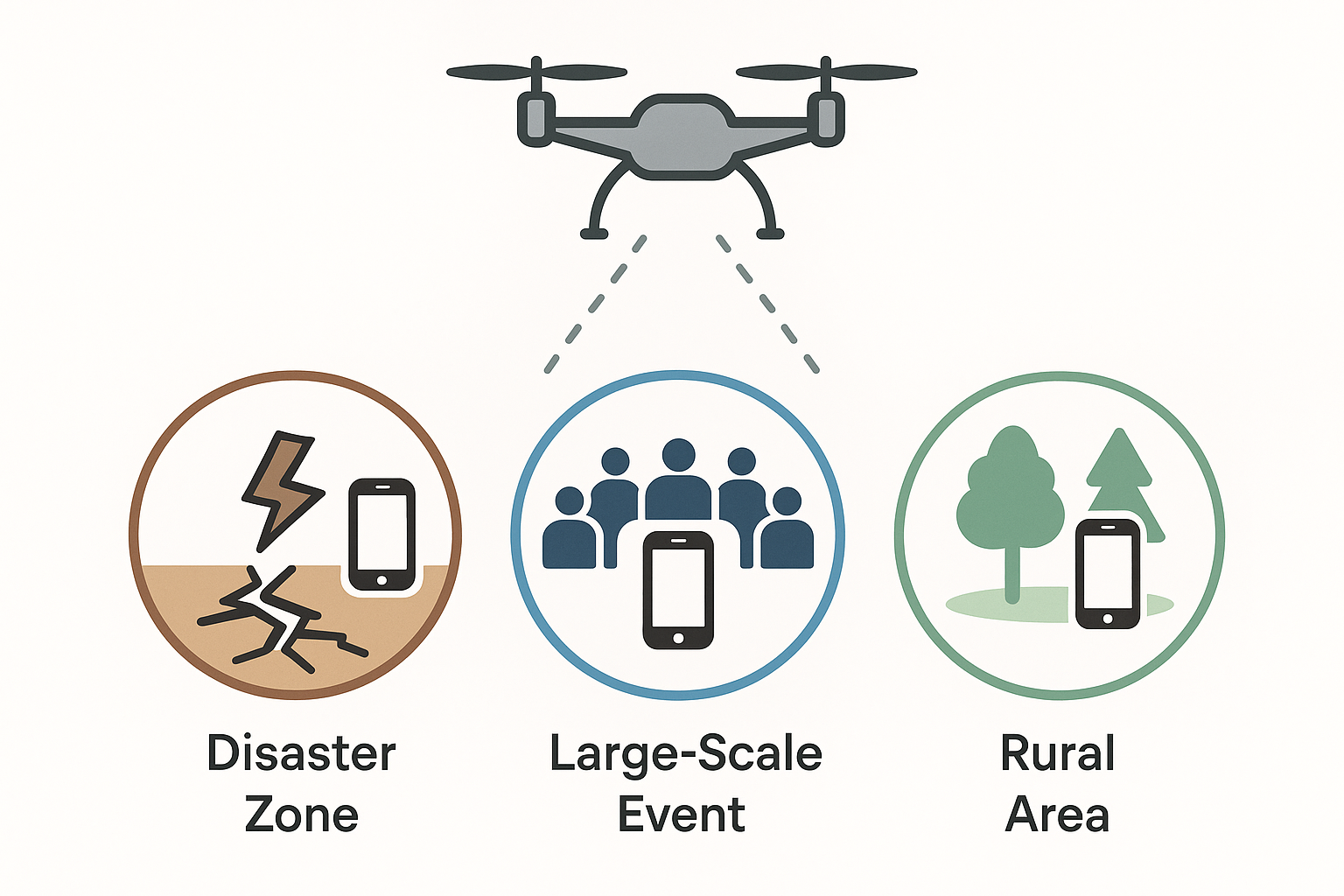}
    \caption{UAV-BS use cases.}
    \label{fig:UAV-BS}
    \vspace{-0.55cm}
\end{figure}

\section{Problem Context and Motivation}
\begin{table*}[t]
\centering
\caption{Comparison of constraints and risks between terrestrial gNBs and UAV-BSs.}
\label{tab:threats}
\begin{tabular}{@{}p{3cm}p{3cm}p{4cm}@{}}
\toprule
\textbf{Category} & \textbf{Terrestrial gNB} & \textbf{UAV-BS} \\ \midrule
Power/Processing & Continuous, high & Battery \& limited CPU \\
\midrule
Backhaul & Wired \& protected & Wireless, jammable \\
\midrule
Physical Access & Secured site & Remotely reachable \\
\midrule
Positioning & Fixed & GNSS-dependent, spoofable \\ \bottomrule
\end{tabular}
\end{table*}

Despite their benefits, UAV-BS come with significant cybersecurity challenges. Some of these are shared with terrestrial base stations, also known as gNBs, or differ fundamentally from them:

\begin{itemize}
\item  {\bf Platform constraints:} UAV-BSs have limited energy, processing, and payload capacity, making it challenging to implement computationally expensive and energy-demanding security mechanisms. Denial-of-service (DoS) attacks targeting the radio access network (RAN), such as radio resource control (RRC) signaling storms \cite{nguyen_rrc_2025}, are expected to have a more severe impact on UAV-BSs than on terrestrial base stations. Battery-draining attacks represent an additional risk specific to UAV-BS \cite{bajracharya_6g_2022}. 

\item {\bf Wireless backhaul vulnerability:} Unlike terrestrial gNBs, which utilize secure wired backhaul links, UAV-BSs rely on wireless feeder links, making them vulnerable to jamming and interception \cite{banafaa_comprehensive_2024}. Jamming can disrupt the connectivity of many user devices \cite{jeong_-rocking_2023,jang_paralyzing_2023}, while interception may leak sensitive control-plane information exchanged between the access and core networks \cite{kim_touching_2019}. 

\item {\bf 	Navigation and positioning risks:} UAVs depend on the Global Navigation Satellite System (GNSS) for flight control. GNSS spoofing can misdirect UAV flight paths, trigger coverage blackouts, or force them into restricted zones \cite{rados_recent_2024}. Such attacks may even cause UAV collisions or border violations, exposing them to capture or destruction. 

\item {\bf Device impersonation:} Adversaries can deploy rogue UAVs to impersonate legitimate gNBs. These fake gNBs can be used for identity catching, location tracking, man-in-the-middle (MitM) attacks, or broadcasting fraudulent emergency alerts \cite{park_why_2023,bitsikas_you_2022}. While these attacks are well-documented in TNs, their exploitation in UAV-based systems remains to be addressed. 

\item {\bf Handover manipulation:} UAV-BSs adjust their positions dynamically to optimize coverage, support user mobility, and balance network loads. The management of user equipment (UE) handovers between TN and NTN, as well as intra-NTN, remains a critical challenge \cite{narmeen_coordinated_2025}. Fake UAV-BSs transmitting at higher power levels can lure into illegitimate handovers, opening the door to MitM and DoS attacks \cite{bitsikas_dont_2021}. 
\end{itemize}
Table~\ref{tab:threats} summarizes UAV-BS and terrestrial gNB constraints and risks. 

Beyond technical challenges, UAV-based systems operate within complex regulatory and operational constraints. UAV operations require permits from general aviation authorities, which enforce altitude restrictions and no-fly zone constraints. Additionally, UAV-BSs must coexist with TN, requiring careful interference management and compliance with regulatory entities.

\section{Attacks and Defenses on UAV-based Cellular Systems}
Securing TN has proven challenging due to the complexity of standards, vendor-specific implementations, backward compatibility requirements, and the presence of unauthenticated broadcast signals \cite{park_why_2023,bitsikas_you_2022,luo_sni5gect_2025} (and references therein). Extending these challenges to 6G NTNs, particularly UAV-BSs, introduces further vulnerabilities related to wireless backhauling and limited power and computational resources, while also creating novel opportunities for new defensive strategies that leverage UAV mobility \cite{banafaa_comprehensive_2024,bajracharya_6g_2022}. In the following, we will divide these attacks into two folds: impersonation attacks using UAVs as rogue base stations and attacks directly targeting UAV-BS. 

\subsection{Impersonating UAV Base Stations}

\subsubsection{Spoofing of Emergency Alerts}
Emergency alerts are one of the most sensitive services offered by mobile operators. They are designed to reach users in a given area with high priority, aiming to notify people about threats such as earthquakes, floods, terrorist attacks, or missing children. Upon reception, loud sounds and vibrations are generated to ensure immediate attention even when phones/tablets are in silent mode. These alerts are delivered through system information blocks (SIBs), which, in current 3GPP implementations, are neither authenticated nor encrypted. This design decision maximizes alert reachability but opens the door to emergency alert spoofing \cite{lee2019your,bitsikas_you_2022}. 

\begin{figure}
    \centering
    \includegraphics[width=0.9\linewidth]{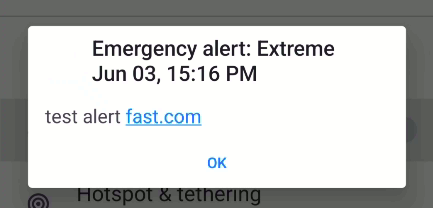}
    \caption{Spoofing of emergency alerts}
    \label{fig:alert}
    \vspace{-0.5cm}
\end{figure}

Authors in \cite{bitsikas_you_2022} demonstrated these attacks for terrestrial gNB using a commercial closed-source solution, lacking the flexibility required for research works. We recently implemented the emergency alert service using the open-source open-air-interface (OAI) project \cite{noauthor_openairinterface_nodate}. The implementation required changes across multiple files to support the creation, scheduling, and transmission of the respective SIB \cite{Abouhasna2025}. Preliminary results indicate the successful reception of alert messages on both Android and iOS phones. We observed that smartphones and tablets parse these alerts, where links, phone numbers, and email addresses are rendered clickable directly from the alert screen (see Fig.~\ref{fig:alert}). This transforms a safety mechanism into a powerful phishing vector. Extending such attacks to UAV-BSs introduces an even greater risk, as UAVs can move across large areas. Moreover, we observed that alert messages can be sent by a rogue gNB and received by the UE even when the core network is offline. Thus, broadcasting fake alerts can be done by the rogue UAV-BS without relying on any core network.  With AI-enabled smartphone assistants, automated exploitation dealing directly with the AI assistant can be imagined. The focus should be on characterizing the user interaction with spoofed alerts and analyzing how different devices parse them. Mitigation techniques are to be investigated and integrated into 6G network standards. One promising approach involves verifying received alerts against governmental alert registries. Another direction is to design integrity checks for SIBs to prevent spoofing \cite{ross_fixing_2024}. 


\subsubsection{Handovers manipulation}
Handover procedures are fundamental in cellular networks. They allow the UE to transition between gNBs without service interruption. However, these procedures rely primarily on signal strength measurements, which makes them vulnerable to impersonation. These measurements are encrypted. Despite this, it has been demonstrated that an attacker setting up a fake gNB, mimicking a legitimate gNB, can exploit vulnerabilities in the handover procedure to cause DoS and MitM attacks, as well as information disclosure. This, in turn, affects both the user and the operator \cite{bitsikas_dont_2021,shaik_new_2019,narmeen_coordinated_2025}. 

In the 6G NTN context, a rogue UAV-BS can maneuver to stay close to targeted UEs, thus making the attack more effective. Indeed, to have a successful attack, the attacker should send their signal in the victim UE's frequency with a higher signal strength than the legitimate one. To counter these attacks, anomaly detection strategies are to be investigated, such as profiling normal handover behavior and flagging deviations caused by rogue UAVs.   

\subsection{Attacks on UAV-BSs}
\subsubsection{DoS via RRC storm attacks}
\begin{figure}
    \centering
    \vspace{-0.2cm}
    \includegraphics[width=0.75\linewidth]{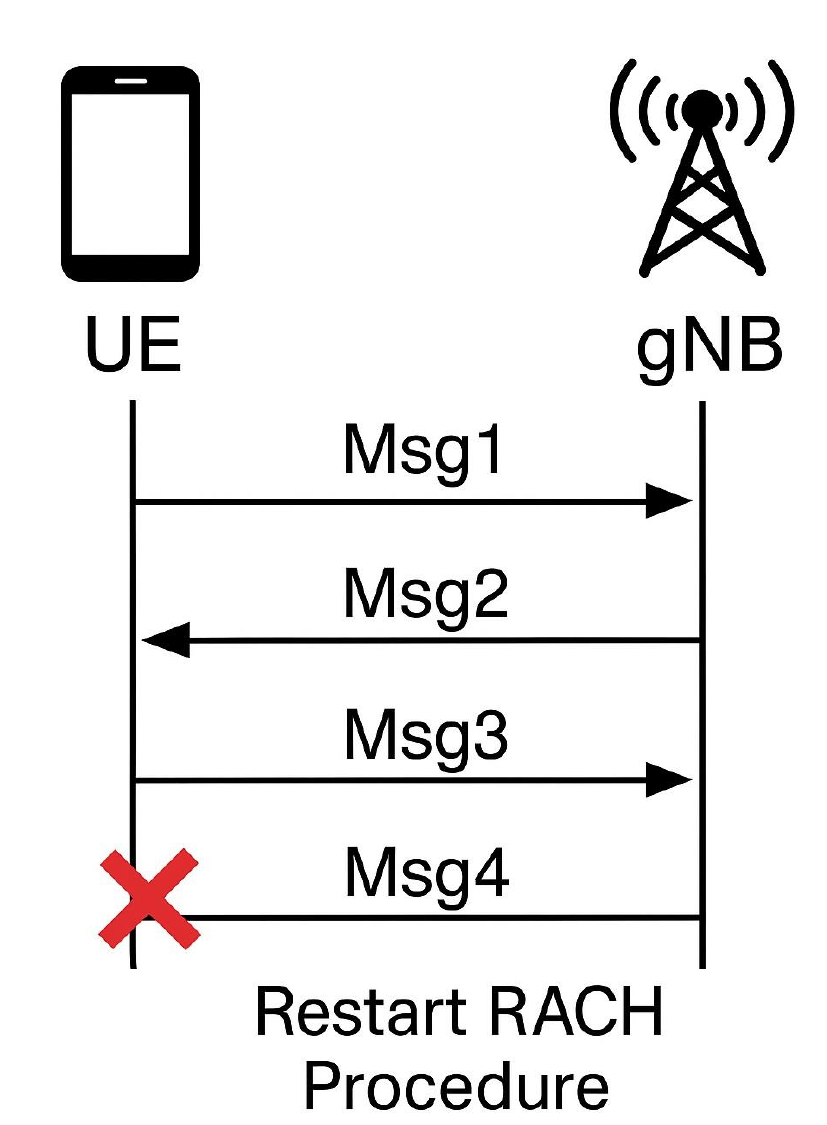}
    \caption{RRC Signaling Storm Attack}
    \label{fig:rrcattack}
    \vspace{-0.5cm}
\end{figure}
The initial attachment and connection procedure, known as the random access channel (RACH) procedure in 5G, is unauthenticated. The gNB allocates resources to the user without receiving and verifying its identity. This makes the RACH procedure vulnerable to repeated connection attempts that exhaust gNB resources, resulting in a DoS. This attack is known as radio resource control (RRC) signaling storm attack \cite{nguyen_rrc_2025}. Fig.~\ref{fig:rrcattack} shows the attack timeline. At Msg4 (RRC setup message) stage, the gNB allocates the resources for the user. At his turn, the attacker continues to restart the RACH procedure, pretending to be a new UE. The gNB will keep allocating resources for the attacker until it is full, resulting in the rejection of normal UE connection attempts.



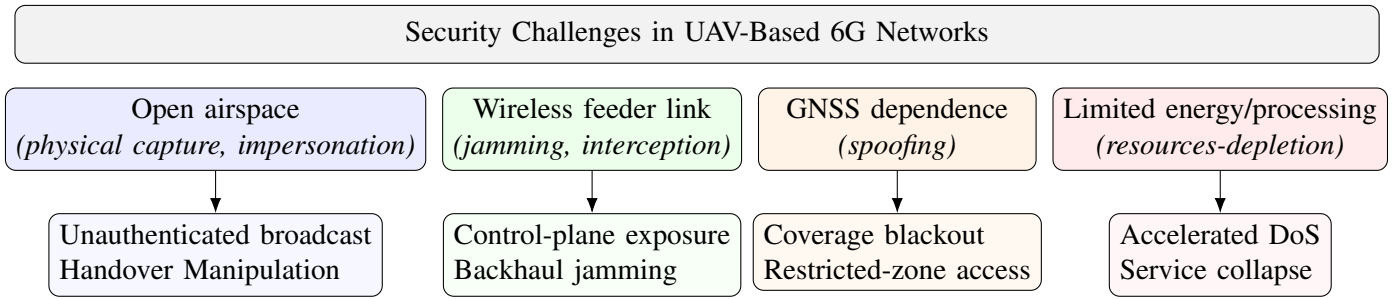
\begin{figure*}[t]
\centering
\begin{tikzpicture}[node distance=6pt]
\node[draw, rounded corners, align=center, minimum width=\linewidth, inner sep=6pt, fill=gray!10] (title) {Security Challenges in UAV-Based 6G Networks};

\node[draw, rounded corners, fill=blue!8, inner sep=4pt, below right=9pt and -520pt of title, align=center] (env) {Open airspace\\ \textit{(physical capture, impersonation)}};
\node[draw, rounded corners, fill=green!8, inner sep=4pt, right=6pt of env, align=center] (bh) {Wireless feeder link\\ \textit{(jamming, interception)}};
\node[draw, rounded corners, fill=orange!10, inner sep=4pt, right=6pt of bh, minimum width= 105pt, align=center] (pos) {GNSS dependence\\ \textit{(spoofing)}};
\node[draw, rounded corners, fill=red!8, inner sep=4pt, right=6pt of pos, align=center] (en) {Limited energy/processing \\ \textit{(resources-depletion)}};

\node[draw, rounded corners, inner sep=4pt, below=16pt of env, align=left, fill=blue!3] (e1) {Unauthenticated broadcast \\Handover Manipulation};
\node[draw, rounded corners, inner sep=4pt, below=16pt of bh, align=left, fill=green!3] (e2) {Control-plane exposure\\Backhaul jamming};
\node[draw, rounded corners, inner sep=4pt, below=16pt of pos, align=left, fill=orange!5] (e3) {Coverage blackout\\Restricted-zone access};
\node[draw, rounded corners, inner sep=4pt, below=16pt of en, align=left, fill=red!3] (e4) {Accelerated DoS\\Service collapse};

\draw[-{Latex[length=2mm]}] (env) -- (e1);
\draw[-{Latex[length=2mm]}] (bh) -- (e2);
\draw[-{Latex[length=2mm]}] (pos) -- (e3);
\draw[-{Latex[length=2mm]}] (en) -- (e4);

\end{tikzpicture}
\caption{Security challenges introduced by UAV-BS: the environment, backhaul, positioning, and energy-computational constraints jointly expand the attack surface.}
\label{fig:taxonomy}
\end{figure*}



UAV-BSs are at a higher risk due to their limited computing and energy budgets. The detection and mitigation of RRC storm attacks in UAV-BS environments is to be explored. Existing work \cite{nguyen_rrc_2025} for terrestrial gNB has demonstrated that comparing the number of received connection requests ($N_r$) to the number of successful attachments ($N_s$) provides a baseline for detection. If $N_r >> N_e$, they assume an attack has been performed. No mitigation technique has been proposed. 

By leveraging the fact that the attacker's location remains constant, while connection requests from authentic users are sent from different places, it is possible to detect the attacker. The physical aspects of the signal can be used to differentiate between the attacker and authentic UEs \cite{hanif_unveiling_2025,saleh_integrated_2025}. Mathematical modeling and/or lightweight classification algorithms can be used. The received power, the angle of arrival (AoA), and the distribution of connection requests are promising elements for this challenge. Furthermore, mitigation techniques can be proposed by discarding requests coming from identified attackers. By leveraging the mobility of UAV-BS, more mitigation techniques are possible, including adaptive repositioning of UAV-BSs and the deployment of additional UAVs to bypass localized attacks. 
\subsubsection{Jamming and GNSS spoofing}
UAV-BSs depend on two wireless techniques that are particularly vulnerable: backhaul links for connectivity and GNSS signals for navigation. Both can be exploited to disrupt UAV operations. Jamming attacks on the backhaul link can cause large-scale DoS. On the other hand, GNSS spoofing can mislead UAVs into incorrect flight paths, create coverage blackouts, or even push them into restricted locations \cite{banafaa_comprehensive_2024,rados_recent_2024,saleh_integrated_2025,kim_touching_2019}. For jamming, multiple jamming categories exist as: constant, reactive, random, and deceptive \cite{manesh_performance_2019}. Furthermore, GNSS spoofing can alter UAV trajectories and positioning.

Defenses are then to be explored. For jamming, mitigation techniques include: 1) beam nulling, where the receiver is deactivated in a specific direction, 2) UAV repositioning, where the UAV-BS changes its position to avoid targeted jamming, and 3) cooperative defense using additional UAV-BSs, or the possibility of using an additional UAV-BS when the main UAV-BS is under jamming. For GNSS spoofing, mitigation techniques include: multi-constellation fusion across GPS, Galileo, and BeiDou, combined with signal power monitoring and angle-of-arrival estimation \cite{rados_recent_2024}. One additional mitigation technique specific to 6G TN-NTN is the possibility of cross-checking the position of UAV-BS with the TN. Backhaul-assisted validation, where UAV positions are cross-checked against those determined by terrestrial gNBs, is a promising approach. \\

Fig.~\ref{fig:taxonomy} summarizes the security challenges introduced by UAV-BS in 6G systems.

\section{Position and Call to Action}
When connectivity extends into the air, terrestrial network assumptions about security must be reconsidered. Securing UAV-based networks requires integrating security as a design principle in NTN standards, cross-disciplinary cooperation between cybersecurity, communications, and aviation communities. A tradeoff between processing performance, permitting efficient security measures, and a lightweight design for better flying capabilities is essential.    

\section{Conclusion}
UAV-based 6G NTNs will be indispensable for emergency connectivity and future smart-city ecosystems. In this paper, we presented the security challenges introduced by UAV-based 6G networks. Mitigation techniques are also discussed. These challenges must be addressed to ensure a secure, fully integrated air-ground 6G system. 

\bibliographystyle{IEEEtran}
\bibliography{IEEEabrv,references} 

\end{document}